\documentclass[aps,pra,twocolumn,showpacs]{revtex4}
\usepackage{amssymb,amsmath,amsfonts}
\usepackage{bbm,wasysym}
\usepackage{graphics}

\newcommand{\fig}[4]{\begin{figure}%
 \setlength{\unitlength}{0.9\columnwidth}%
 \begin{picture}(1,#2)
  \resizebox{\unitlength}{!}{\includegraphics{#1}}
  #3
\end{picture}%
 \caption{\label{#1}#4}%
\end{figure}FIG.~\ref{#1}}

\newcommand{\figNL}[3]{\begin{figure}%
 \setlength{\unitlength}{0.9\columnwidth}%
 \begin{picture}(1,#2)
  \resizebox{\unitlength}{!}{\includegraphics{#1}}
  #3
\end{picture}\end{figure}}

\newtheorem{proposition}{Proposition}
\newcommand{\qed}{\hfill$\blacksquare$}
\newcommand{\mx}[1]{\quad \mbox{#1} \quad}
\newcommand{\ds}{\displaystyle}
\newcommand{\cs}{\textstyle}
\newcommand{\sfrac}[2]{{\textstyle \frac{#1}{#2}}}
\newcommand{\half}{{\textstyle \frac 1 2}}
\newcommand{\coloneqq}{:=}
\newcommand{\bra}[1]{{\langle #1 |}}
\newcommand{\ket}[1]{{| #1 \rangle}}
\newcommand{\bracket}[2]{{\langle #1 | #2 \rangle}}
\newcommand{\pr}[1]{{\ket{#1}\!\bra{#1}}}
\newcommand{\id}{{\mathbbm 1}}
\newcommand{\Kern}[1]{{{\mathcal K}_{#1}}}
\newcommand{\Supp}[1]{{{\mathcal S}_{#1}}}
\newcommand{\Hilbert}{{\mathcal H}}
\DeclareMathOperator{\Span}{span}
\DeclareMathOperator{\tr}{tr}
\newcommand{\NTS}{\mathcal K^\cap}

\begin{document}
 \title{On the generalization of quantum state comparison}
 \author{M. Kleinmann}
 \email{kleinmann@thphy.uni-duesseldorf.de}
 \author{H. Kampermann}
 \author{D. Bru\ss}
 \affiliation{Institut f\"ur Theoretische Physik III,
              Heinrich-Heine-Universit\"at
              D\"usseldorf, D-40225 D\"usseldorf, Germany}
 \date{\today}

\begin{abstract}
We investigate the unambiguous comparison of quantum states in a scenario that
is more general than the one that was originally suggested by Barnett {\em et
al.} First, we find the optimal solution for the comparison of two states taken
from a set of two pure states with arbitrary {\em a priori} probabilities. We
show that the optimal coherent measurement is always superior to the optimal
incoherent measurement. Second, we develop a strategy for the comparison of two
states from a set of $N$ pure states, and find an optimal solution for some
parameter range when $N=3$. In both cases we use the reduction method for the
corresponding problem of mixed state discrimination, as introduced by Raynal
{\em et al.}, which reduces the problem to the discrimination of two pure
states only for $N=2$. Finally, we provide a necessary and sufficient condition
for unambiguous comparison of mixed states to be possible.
\end{abstract}

\maketitle

\section{Introduction}
The laws of quantum mechanics do not allow the perfect discrimination of two
non-orthogonal quantum states $\ket{\psi_1}$ and $\ket{\psi_2}$. Consequently,
given a set of non-orthogonal states $\{\ket{\psi_1},\ket{\psi_2}\}$, it is
also impossible to find out with probability one whether two quantum states,
drawn from this set, are identical (namely, the total state is either
$\ket{\psi_1\psi_1}$ or $\ket{\psi_2\psi_2}$) or different (i.e. the total
state is either $\ket{\psi_1\psi_2}$ or $\ket{\psi_2\psi_1}$). What is the
optimal probability of success, when no errors are allowed? This problem has
been introduced by Barnett, Chefles and Jex \cite{Barnett:2003PLA} and is
called unambiguous quantum state comparison. It has been solved for the case
that the {\em a priori} probabilities for the two ensemble states are equal
\cite{Barnett:2003PLA}. The task of determining whether $C$ given states taken
from a set of $N$ pure states with equal {\em a priori} probabilities are
identical or not has been investigated in
\cite{Chefles:2004JPA,Chefles:2004JMO}.

In this article, we consider the most general case of unambiguous state
comparison, also admitting mixed states. We provide sufficient and necessary
conditions, for which this task can succeed. Furthermore, the comparison of two
states drawn from a set of $N$ pure states with arbitrary {\em a priori}
probabilities is investigated, and an optimal solution is found for the case
$N=2$, as well as for a range of parameters in the case $N=3$, using the
reduction techniques for mixed state discrimination developed in
\cite{Raynal:2003PRA}. This method is also applied for general N.  We address
the question of how much can be gained in the optimal coherent strategy (i.e.
with global measurements on the two given states), as compared to the best
incoherent strategy (i.e. consecutive measurements).

Our paper is organized as follows: in section~\ref{s16313}, we define the most
general state comparison problem, and explain the connection to mixed state
discrimination. In section~\ref{s32551}, we find the optimal solution for
comparing two states, drawn from a set of two states. In section~\ref{s7931},
we develop the formalism for the comparison of two out of $N$ states, and apply
it to $N=3$. In section~\ref{s19384} we derive sufficient and necessary
conditions for the general task of mixed state comparison to be successful,
before concluding in section~\ref{s29399}.

\section{General State Comparison}\label{s16313}
Let us define the task of state comparison in the most general way: {\em Given
$C$ quantum states of arbitrary dimension, each of them taken from a set of $N$
possible (in general mixed) quantum states $\{\pi_1, \dotsc, \pi_N\}$ that
occur with corresponding {\em a priori} probabilities $\{q_1, \dotsc, q_N\}$.
Unambiguous state comparison ``$C$ out of $N$'' is performed by doing a
measurement, which allows with probability $P$ to decide without doubt whether
all $C$ states are equal, or whether at least one of them is different. The
best possible probability of success $P_\mathrm{opt}$ is reached in optimal
state comparison.}

A measurement is most generally described as a positive operator-valued
measurement (POVM), i.e. a decomposition of the identity operator into a set of
$n$ positive operators \cite{Peres:Book}
\begin{equation}
 F_1, \dotsc, F_n\ge 0, \mx{satisfying} \sum_i F_i= \id.
\end{equation}
The probability for a system in a state $\varrho_k$ to yield the outcome
corresponding to $F_i$ is given by $p_k\tr(F_i\varrho_k)$, where $p_k$ is the
{\em a priori} probability for the system being in state $\varrho_k$. For the
task of unambiguous state comparison, we need at least two measurements $F_a$
and $F_b$, having vanishing probabilities in the case where the total state is
composed of different or equal states, respectively. This means, that for all
$(p_k, \varrho_k) \in \{ (q_{i_1}\dotsm q_{i_C}, \pi_{i_1}\otimes \dotsm
\otimes \pi_{i_C}) \mid i_1, \dotsc, i_C \in \{1, \dots, N\} \}$ we demand
\begin{subequations}\label{e6923}
\begin{eqnarray}
 p_k\tr(F_a \varrho_k) > 0 &\Leftrightarrow& \exists m\colon \varrho_k=
  {\pi_{m}}^{\otimes C} \label{e11971},\\
 p_k\tr(F_b \varrho_k) > 0 &\Leftrightarrow& \nexists m\colon \varrho_k=
  {\pi_{m}}^{\otimes C}.
\end{eqnarray}
\end{subequations}
However, measurements which satisfy this defining property will in general not
sum up to the identity, thus admitting the inconclusive measurement $F_?= \id-
F_a- F_b$, which has to be a positive operator. In order to find an {\em
optimal} solution to the problem, one has to minimize the probability for the
inconclusive answer $\sum_k p_k\tr({F_?\varrho_k})$, or equivalently maximize
the rate of success given by
\begin{equation}\label{e27816}
 P= \sum_k p_k\tr((F_a+ F_b)\varrho_k).
\end{equation}

The problem of finding the optimal measurement for state comparison can be
addressed by considering the optimal solution of a related problem, namely
unambiguous state discrimination. Here, two states $\rho_a$ and $\rho_b$ have
to be distinguished without error, but admitting an inconclusive answer. In
order to see the connection between the two tasks, consider the mixed states %
\begin{subequations}\label{e20410}
\begin{eqnarray}
 \rho_a&=& \frac{1}{\eta_a}\sum_i (q_i\pi_i)^{\otimes C}, \\
 \rho_b&=& \frac{1}{\eta_b}\left(\sum_i q_i\pi_i\right)^{\otimes C}
           -\frac{\eta_a}{\eta_b}\rho_a,
\end{eqnarray}
with {\em a priori} probabilities
\begin{equation}
 \eta_a= \sum_i {q_i}^C \mx{and} \eta_b=1-\eta_a.
\end{equation}
\end{subequations}
Now, a POVM, which satisfies (\ref{e6923}) also has
\begin{equation}\label{e27918}
 F_a\rho_b = 0 \mx{and} F_b\rho_a= 0,
\end{equation}
and furthermore the probability of success (\ref{e27816}) which has to be
optimized can be rewritten as
\begin{equation}
 P= \eta_a \tr(F_a \rho_a)+ \eta_b \tr(F_b \rho_b).
\end{equation}
These equations are characteristic for unambiguous state discrimination. Thus
an optimal solution to the problem of unambiguous discrimination (UD) of
$\rho_a$ and $\rho_b$, which in addition satisfies (\ref{e6923}), is also the
optimal solution to the related problem of unambiguous state comparison. The
task of optimal UD of mixed states has been studied in the literature
\cite{Hillery:2002PRA,Rudolph:2003PRA,Raynal:2003PRA,Zhang:2004QPh,%
Raynal:2005QPh}.

\section{State Comparison ``two out of two''}\label{s32551}
We first consider explicitly the most simple case of state comparison, namely
``two out of two'' with the states subject to comparison being pure states
$\ket{\psi_1}$ and $\ket{\psi_2}$, both of which are vectors in a Hilbert space
of any dimension. The two states may appear with arbitrary (but non-vanishing)
{\em a priori} probabilities $q_1$ and $q_2$. The trivial cases, where both
states are co-linear or orthogonal are not considered. Without loss of
generality the phase between the two states can be chosen to be real, so that
their overlap is determined by their relative angle $\vartheta$,
\begin{equation}
 \cos\vartheta\coloneqq \bracket{\psi_1}{\psi_2}\in \left]0, 1\right[.
\end{equation}

We consider the related UD problem of the corresponding mixed states, which are
according to the equations~(\ref{e20410}) given by
\begin{subequations}
\begin{eqnarray}
 \rho_a&=& \sfrac{1}{\eta_a} (q_1^2\, \pr{\psi_1\psi_1}+
                              q_2^2\, \pr{\psi_2\psi_2}),\\
 \rho_b&=& \half (\pr{\psi_1\psi_2}+ \pr{\psi_2\psi_1}),
\end{eqnarray}
appearing with {\em a priori} probabilities
\begin{equation}
 \eta_a= q_1^2+ q_2^2 \mx{and} \eta_b= 2 q_1 q_2.
\end{equation}
\end{subequations}
Note, that $\eta_a\ge \eta_b$ always holds. In what follows, we construct an
optimal solution of this related UD problem and then show that the POVM of this
solution satisfies (\ref{e6923}), thus providing an optimal solution of the
unambiguous state comparison task.

\subsection{Reduction to the Non-Trivial Subspace}
It has been shown by Raynal, L\"utkenhaus and van Enk \cite{Raynal:2003PRA}
that the optimal UD of mixed states can be reduced to a subspace of the Hilbert
space in such a way, that the relevant density matrices, acting on the reduced
space, have equal rank and their kernels form non-orthogonal subspaces, the
intersection of which is zero. This is achieved in two reduction steps: In the
{\em first reduction step}, the Hilbert space is reduced to its non-trivial
part, removing that part of the Hilbert space, where no UD is possible at all.
We will denote this reduced space as $\Hilbert$. It is given by the particular
space, where
\begin{equation}
 \Supp{\rho_a}\cap \Supp{\rho_b}= 0 \mx{and} \Kern{\rho_a}\cap \Kern{\rho_b}= 0
\end{equation}
holds. Here $\Kern{\rho}$ is the kernel of $\rho$ and $\Supp{\rho}$ its
support, defined as the ortho-complement to the kernel \footnote{Since a
density matrix is Hermitian, in particular the support of a density matrix is
identical to its range.}. Thus, $\Hilbert$ contains only the direct sum of the
support of $\rho_a$ and $\rho_b$, i.e. $\Hilbert= \Supp{\rho_a}\oplus
\Supp{\rho_b}$.

For our system, we have
\begin{subequations}
\begin{eqnarray}
 \Supp{\rho_a}&=& \Span( \ket{\psi_1\psi_1},\ket{\psi_2\psi_2}),\\
 \Supp{\rho_b}&=& \Span( \ket{\psi_1\psi_2},\ket{\psi_2\psi_1}),
\end{eqnarray}
\end{subequations}
which already satisfy $\Supp{\rho_a}\cap \Supp{\rho_b}= \{0\}$ due to the
linear independence of $\ket{\psi_1}$ and $\ket{\psi_2}$. For the further
calculation it is convenient to rewrite both supports in an appropriate basis
of $\Hilbert$. Therefore consider complementary normalized vectors
$\ket{\overline\psi_1}, \ket{\overline\psi_2} \in \Span(\ket{\psi_1},
\ket{\psi_2})$, which are in the same plane as $\ket{\psi_1}$ and
$\ket{\psi_2}$, but orthogonal to the corresponding vector, i.e.
$\ket{\overline\psi_1} \perp \ket{\psi_1}$ and $\ket{\overline\psi_2} \perp
\ket{\psi_2}$. Then, an orthonormal basis of $\Hilbert$ is given by
\begin{subequations}
\begin{eqnarray}\label{e10447}
 \ket{e_{1,2}}&=&\sfrac{1}{\sqrt 2\,n_\pm}
                  (\ket{\psi_1\psi_1}\pm
                   \ket{\psi_2\psi_2}),\\\label{e10448}
 \ket{e_{3,4}}&=&\sfrac{1}{\sqrt 2\,n_\pm}
                  (\ket{\overline\psi_1\overline\psi_2}\pm
                   \ket{\overline\psi_2\overline\psi_1}),
\end{eqnarray}
\end{subequations}
with $ n_\pm= \sqrt{1\pm \cos^2\vartheta}$. In equation (\ref{e10447}), the
$+$($-$)-sign refers to the index $1$($2$) and in (\ref{e10448}) to $3$($4$)
respectively.

By this choice, one immediately has $\Kern{\rho_a}= \Span(\ket{e_3},
\ket{e_4})$ and $\ket{e_2}\in \Kern{\rho_b}$. Let us denote by $\mathrm P_+=
\pr{e_1}+ \pr{e_3}$ ($\mathrm P_-= \pr{e_2}+ \pr{e_4}$) the projector onto that
subspace, which is symmetric (antisymmetric) under exchanging $\ket{\psi_1}$
and $\ket{\psi_2}$. Then, due to $\ket{\psi_1\psi_2}\in \Supp{\rho_b}$,
\begin{equation}
 \ket{\gamma}\coloneqq
  \sfrac{\sqrt 2}{n_+}\, \mathrm P_+ \ket{\psi_1\psi_2}=
  \sfrac{\sqrt 2}{n_+}\, \mathrm P_+\ket{\psi_2\psi_1} \in \Supp{\rho_b}
\end{equation}
must hold, where $\ket{\gamma}$ is normalized and has the components
\begin{equation}
 |\bracket{e_1}{\gamma}|= \sfrac{2 \cos\vartheta}{n^2_+} \mx{and}
 |\bracket{e_3}{\gamma}|= \sfrac{\sin^2\vartheta}{n^2_+}.
\end{equation}
Since $\mathrm P_-+ \mathrm P_+= \id_\Hilbert$, the second spanning vector of
$\Supp{\rho_b}$ has to be $\mathrm P_- \ket{\psi_1\psi_2}= -\mathrm
P_-\ket{\psi_2\psi_1}$. This vector, however, cannot have any component in
direction of $\ket{e_2} \in \Kern{\rho_b}$ and therefore has to be parallel to
$\ket{e_4}$. Thus, we finally write the non-trivial Hilbert space $\Hilbert$ as
\begin{equation}
 \Hilbert= \Supp{\rho_a}\oplus \Supp{\rho_b}
    \equiv \Span(\ket{e_1},\ket{e_2})\oplus \Span(\ket{\gamma},\ket{e_4}).
\end{equation}
Due to the particular choice of basis, we further find $\Kern{\rho_b}=
\Span(\ket{\gamma^\perp},\ket{e_2})$, where
$\ket{\gamma^\perp}$ is a normalized vector satisfying
$\ket{\gamma^\perp}\perp\ket{\gamma}$ and $\mathrm P_-\ket{\gamma^\perp}=
\mathrm P_-\ket{\gamma}\equiv 0$.

\subsection{Optimal Solution}
In the {\em second reduction step} shown in \cite{Raynal:2003PRA}, one reduces
the space by those parts, which allow perfect UD. These parts are given by
\begin{equation}
 \NTS_a\coloneqq \Kern{\rho_a}\cap \Supp{\rho_b} \mx{and}
 \NTS_b\coloneqq \Kern{\rho_b}\cap \Supp{\rho_a}.
\end{equation}
The Hilbert space $\Hilbert$ can then be decomposed into
\begin{equation}
 \Hilbert= \Hilbert'\oplus \NTS_a\oplus \NTS_b,
\end{equation}
where $\Hilbert'$ is conveniently chosen to be the ortho-complement of
$\NTS_a\oplus \NTS_b$.  Denoting by $\mathrm P_{\Hilbert'}$ the projector onto
$\Hilbert'$, and further writing $\zeta_a$, $\zeta_b$ for appropriate
normalization constants, the density matrices
\begin{equation}
 \rho_a'= \sfrac{1}{\zeta_a} \mathrm P_{\Hilbert'} \rho_a \mathrm P_{\Hilbert'}
  \mx{and}
 \rho_b'= \sfrac{1}{\zeta_b} \mathrm P_{\Hilbert'} \rho_b \mathrm P_{\Hilbert'}
\end{equation}
are states acting on $\Hilbert'$ and having {\em a priori} probabilities
\begin{equation}
 \eta_a'= \frac{\eta_a\zeta_a}{\zeta} \mx{and}
\eta_b'=1-\eta_a',
\end{equation}
where $\zeta\coloneqq \zeta_a\eta_a+\zeta_b\eta_b$. Suppose that $P'$ is the
optimal rate of success for this reduced problem. Then the optimal rate of
success of the complete problem was shown \cite{Raynal:2003PRA} to be
\begin{equation}
 P_\mathrm{opt}= 1-(1-P')\zeta.
\end{equation}

In our basis, we immediately find
\begin{subequations}
\begin{eqnarray}
 \NTS_a&=& \Span(\ket{e_3},\ket{e_4})\cap \Supp{\rho_b}
          =\Span(\ket{e_4}),\\
 \NTS_b&=& \Span(\ket{\gamma^\perp},\ket{e_2})\cap \Supp{\rho_a}
          =\Span(\ket{e_2}),
\end{eqnarray}
\end{subequations}
since $\ket{\gamma} \nparallel \ket{e_3}$ and $\ket{\gamma^\perp} \nparallel
\ket{e_1}$ holds.

Now the optimization problem can be reduced to $\Hilbert'=
\Span(\ket{e_1},\ket{e_3})$. Since the remaining problem is two-dimensional, it
can be considered as the well-known discrimination of pure states. Indeed, the
problem reduces to the UD of
\begin{subequations}
\begin{eqnarray}
 \rho'_a&=& \sfrac{1}{\zeta_a} \mathrm P_+\rho_a\mathrm P_+= \pr{e_1}, \\
 \rho'_b&=& \sfrac{1}{\zeta_b} \mathrm P_+\rho_b\mathrm P_+= \pr{\gamma}.
\end{eqnarray}
\end{subequations}
Calculating the normalization factors $\zeta_a= \tr(P_+ \rho_a)$ and $\zeta_b=
\tr(P_+ \rho_b)$, one obtains $\zeta_a= \zeta_b= \zeta= \half n_+^2$ and thus
the {\em a priori} probabilities of the reduced problem remain unchanged,
$\eta'_a\equiv \eta_a$ and $\eta'_b\equiv \eta_b$. Jaeger and Shimony have
derived \cite{Jaeger:1995PLA} the optimal UD of two pure states with an
unbalanced probability distribution. Using their result for the discrimination
between $\ket{e_1}$ and $\ket{\gamma}$, the optimal rate of success for UD of
$\rho_a$ and $\rho_b$ calculates to
\begin{equation}\label{e26445}
 P_\mathrm{opt}=\left\{\begin{array}{lc}
  \ds 1-2\sqrt{\eta_a\eta_b}\cos\vartheta & \quad\mbox{if (\ref{e26601})}\\
  \ds \sfrac{n_-^2}{n_+^2}
      \left(1-\sfrac{\eta_b}{2}\sin^2\vartheta\right)&\quad\mbox{else,}
 \end{array}\right.
\end{equation}
where (\ref{e26601}) is the condition
\begin{equation}\tag{$\ast$}\label{e26601}
  \cos\vartheta< \sqrt{\frac{\eta_a}{\eta_b}}
   \left(1- \sqrt{\frac{\eta_a-\eta_b}{\eta_a}}\right).
\end{equation}

Further, the optimal POVM of the reduced problem is given by
\begin{equation}
 F'_a= \alpha \pr{\gamma^\perp} \mx{and} F'_b=\beta \pr{e_3}.
\end{equation}
In the region, where (\ref{e26601}) holds,
\begin{subequations}
\begin{eqnarray}
 \alpha&=& \frac{1-
  \sqrt{\frac{\eta_b}{\eta_a}}
    |\bracket{e_1}{\gamma}|}{|\bracket{e_3}{\gamma}|^2},\\
 \beta&=& \frac{1-
  \sqrt{\frac{\eta_a}{\eta_b}}
    |\bracket{e_1}{\gamma}|}{|\bracket{e_3}{\gamma}|^2},
\end{eqnarray}
\end{subequations}
and $\alpha=  1$, $\beta= 0$ elsewhere. The optimal measurement of the full
problem is then given by
\begin{equation}
 F_a= F'_a+ \mathrm P_{\NTS_b} \mx{and} F_b= F'_b+ \mathrm P_{\NTS_a},
\end{equation}
where $\mathrm P_{\NTS_b}\equiv \pr{e_2}$ and $\mathrm P_{\NTS_a}\equiv
\pr{e_4}$. The fact that the projectors $\pr{e_2}$ and $\pr{e_4}$ have to be
part of the optimal POVMs $F_a$ and $F_b$, respectively, was already obvious
from the structure of the kernels and supports, since $\ket{e_2}$ and
$\ket{e_4}$ are orthogonal and part of either $\Supp{\rho_a}$ or
$\Supp{\rho_b}$.

Now one easily verifies, that condition~(\ref{e6923}) holds for this
measurement, by noting that $|\bracket{\psi_1\psi_1}{e_2}|^2=
|\bracket{\psi_2\psi_2}{e_2}|^2>0$ and $|\bracket{\psi_1\psi_2}{e_4}|^2=
|\bracket{\psi_2\psi_1}{e_4}|^2>0$. Thus we have found the optimal solution for
unambiguous two-dimensional state comparison. Furthermore, as we discuss in the
following, this solution is {\em always} better then a separable measurement on
both states, which becomes manifest by the fact, that $F_a$ and $F_b$ are not
separable, i.e. the partial transpose fails to be positive semidefinite.

\subsection{Discussion}
In the literature, an optimal solution for the problem of state comparison has
only been found for the case of equal probabilities. Barnett, Chefles and Jex
\cite{Barnett:2003PLA} showed, that in this case the optimal rate of success is
given by $P= 1- \cos\vartheta$, which is our result for $q_1= q_2= \half$. This
particular result was also obtained by Rudolph, Spekkens and Turner
\cite{Rudolph:2003PRA}, by providing a general upper and lower bound for the
rate of success of an UD of mixed states.  Their upper bound matches our result
only in situations, where (\ref{e26601}) holds. On the other hand, their lower
bound turns out to match our optimal result for all parameters and thus our
calculation has proven, that their lower bound is indeed optimal for the UD of
$\rho_a$ and $\rho_b$.

Let us compare our result with the na\"{\i}ve incoherent strategy, where both
states are measured consecutively. The straightforward approach of the optimal
POVM $\{\widetilde F_1, \widetilde F_2, \widetilde F_?\}$ for unambiguous
discrimination between $\ket{\psi_1}$ and $\ket{\psi_2}$, leads to
\begin{subequations}\label{e20301}
\begin{eqnarray}
 F^\mathrm{sep}_a&=& \widetilde F_1\otimes \widetilde F_1+
                       \widetilde F_2\otimes \widetilde F_2, \\
 F^\mathrm{sep}_b&=& \widetilde F_1\otimes \widetilde F_2+
                       \widetilde F_2\otimes \widetilde F_1.
\end{eqnarray}
\end{subequations}
This na\"{\i}ve method is indeed the {\em optimal} separable measurement, as
shown in appendix~\ref{a348}. It has a rate of success given by the square of
the success probability for unambiguous discrimination of $\ket{\psi_1}$ and
$\ket{\psi_2}$, i.e. \cite{Jaeger:1995PLA}
\begin{equation}
 P_\mathrm{sep}= \left\{ \begin{array}{ll}
  (1-2\sqrt{q_1q_2}\cos\vartheta)^2 & \mx{if (\ref{e31616})} \\
  q_\mathrm{max}^2\sin^4\vartheta & \mx{else,}
 \end{array}\right.
\end{equation}
where $q_\mathrm{max}$ is the maximum of $q_1$ and $q_2$, and (\ref{e31616}) is
the condition
\begin{equation} \tag{$\ast\ast$}\label{e31616}
 \cos\vartheta< \sqrt{\frac{1-q_\mathrm{max}}{q_\mathrm{max}}}.
\end{equation}
In \fig{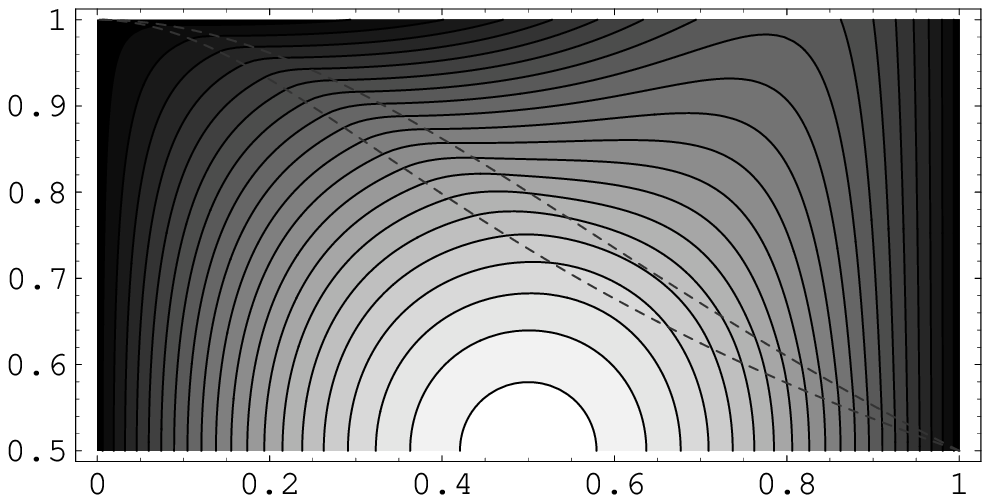}{0.5}{
 \put(-1.04,0.27){\rotatebox{90.0}{\makebox(0,0){$q_1$}}}
 \put(-0.46,-0.02){\makebox(0,0){$\cos\vartheta$}}
}{Contour plot of the gain $P_\mathrm{opt}- P_\mathrm{sep}$, where higher gain
corresponds to brighter shade. White stands for a gain value of $0.25$, black
for a value of $0.0125$, and each contour line corresponds to a step of
$0.0125$. The dashed lines divide the set of parameters into regions where both
(\ref{e26601}) and (\ref{e31616}) hold (lower left), neither of both condition
holds (top right) and (\ref{e31616}) holds, but (\ref{e26601}) does not
(remaining small stripe).} we show the gain $P_\mathrm{opt}- P_\mathrm{sep}$,
which of course is always positive or zero. This gain has its absolute maximum
of $\frac 1 4$ at $q_1= \half$ and $\vartheta= \sfrac \pi 3$. While for fixed
angles the maximum gain is always at $q_1= \half$, one finds for fixed {\em a
priori} probabilities, that at some regions there are two maxima. The maximum
in low values of $\cos\vartheta$ appears, where (\ref{e31616}) holds without
having (\ref{e26601}) satisfied. Also note, that the gain function is
asymmetric in $\cos\vartheta$, while it is symmetric in $q_1$. In
\figNL{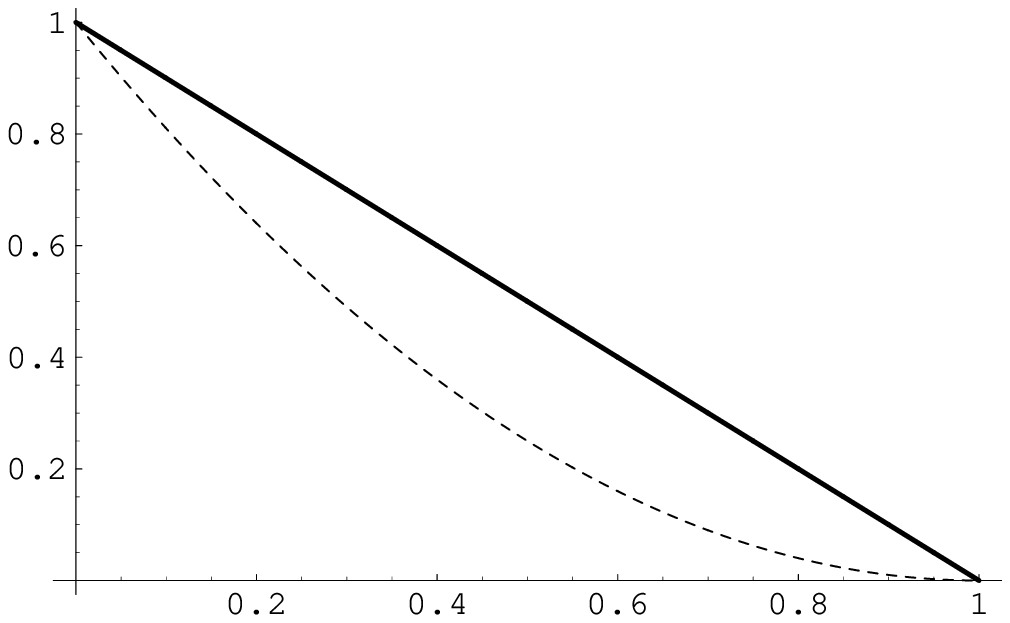}{0.61804}{
 \put(-1.04,0.34){\rotatebox{90.0}{\makebox(0,0){$P$}}}
 \put(-0.46,-0.02){\makebox(0,0){$\cos\vartheta$}}
 \put(-0.30,0.42){\makebox(0,0){$q_1=\half$}}
}\fig{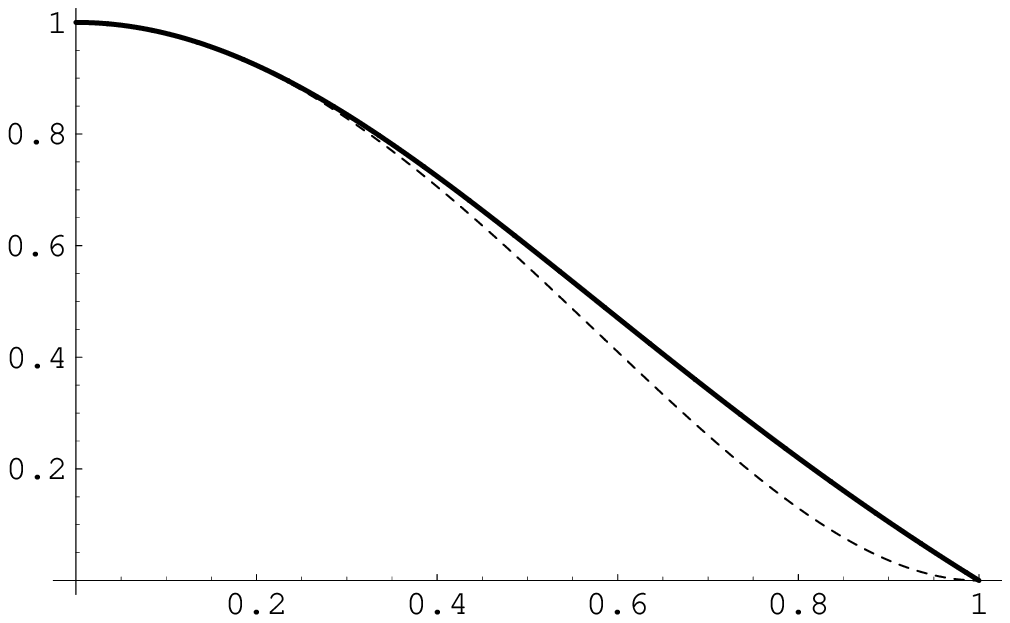}{0.61804}{
 \put(-1.04,0.34){\rotatebox{90.0}{\makebox(0,0){$P$}}}
 \put(-0.46,-0.02){\makebox(0,0){$\cos\vartheta$}}
 \put(-0.30,0.42){\makebox(0,0){$q_1\rightarrow 1$}}
}{Rate of success for state comparison ``two out of two'' with $q_1= \half$
(upper graph) and $q_1\rightarrow 1$ (lower graph). The solid line is the
optimal result, and the dashed line corresponds to the best separable
measurement.}, the gain of the coherent versus the incoherent strategy is
illustrated for the parameters $q_1=\half$ and $q_1\rightarrow 1$.

\section{State Comparison ``two out of $N$''}\label{s7931}
Next, we investigate the problem of unambiguous state comparison ``two out of
$N$'' for pure states. As shown by Chefles {\em et al.} \cite{Chefles:2004JPA}
for equal probabilities and in section~\ref{s19384} for arbitrary
probabilities, this can only work if all $N$ states are linearly independent,
thus spanning an $N$-dimensional Hilbert space. Again this unambiguous state
comparison is related to the UD of
\begin{subequations}
\begin{eqnarray}
 \rho_a&=& \sfrac{1}{\eta_a} \sum_i^N q_i^2\pr{\psi_i\psi_i}, \\
 \rho_b&=& \sfrac{1}{\eta_b} \sum_{i\ne j}^N q_iq_j\pr{\psi_i\psi_j},
\end{eqnarray}
having {\em a priori} probabilities
\begin{equation}
 \cs \eta_a=\sum q_i^2, \quad \eta_b= \sum_{i\ne j} q_iq_j.
\end{equation}
\end{subequations}
We immediately obtain
\begin{subequations}
\begin{eqnarray}
 \Supp{\rho_a}&=&\bigoplus_i \Span(\ket{\psi_i\psi_i}),\\
 \Supp{\rho_b}&=&\bigoplus_{i\ne j}\Span(\ket{\psi_i\psi_j}) \nonumber\\
 &=&\bigoplus_{i> j} \Span(\ket{\psi_i\psi_j}\pm\ket{\psi_j\psi_i}).
\end{eqnarray}
\end{subequations}
Due to linear independence $\Supp{\rho_a}\cap \Supp{\rho_b}= \{0\}$ holds and
thus the first reduction step yields $\Hilbert= \Supp{\rho_a}\oplus
\Supp{\rho_b}$. Note, that the dimension of $\Supp{\rho_a}$ is now in general
much smaller then the one of $\Supp{\rho_b}$, because $\dim \Supp{\rho_a}= N$
while $\dim \Supp{\rho_b}= N^2-N$. In what follows we show in a constructive
way, that the $N$-dimensional state comparison in general is related to such an
UD of mixed states, which cannot be reduced to UD of pure states.

The second reduction step can be performed as follows. The antisymmetric
subspace $\Hilbert^-= \bigoplus_{i> j} \Span( \ket{\psi_i\psi_j}-
\ket{\psi_j\psi_i})$ is part of $\NTS_a\equiv \Kern{\rho_a}\cap \Supp{\rho_b}$,
since
\begin{equation}
 \Supp{\rho_a} \perp \Hilbert^- \mx{and} \Supp{\rho_b} \supset \Hilbert^-.
\end{equation}
Further, $\Supp{\rho_a}$ is part of the symmetric subspace $\Hilbert^+=
\bigoplus_{i\ge j} \Span(\ket{\psi_i\psi_j}+ \ket{\psi_j\psi_i})$ and thus, due
to $\Hilbert^-\perp \Hilbert^+$, we have the orthogonal decomposition
\begin{equation}
 \NTS_a= {\NTS_a}^- \oplus {\NTS_b}^+,
\end{equation}
with ${\NTS_a}^-\coloneqq \Hilbert^-$ and ${\NTS_a}^+\coloneqq \Hilbert^+ \cap
\Kern{\rho_a}$. In order to obtain ${\NTS_a}^+$, let $C_{ij} \coloneqq
\bracket{\psi_i}{\psi_j}$ be the Hermitian overlap matrix and $A_{ij}$ be a
lower triangular coefficient matrix. Then ${\NTS_a}^+$ is given by all vectors
$\sum_{i>j}A_{ij} (\ket{\psi_i\psi_j}+ \ket{\psi_j\psi_i})$, which satisfy
\begin{eqnarray}
 &\forall k& \bra{\psi_k\psi_k}
             \sum_{i>j}A_{ij}(\ket{\psi_i\psi_j}+ \ket{\psi_j\psi_i})= 0
               \nonumber\\
 \Leftrightarrow
 &\forall k& \sum_{i>j} C_{ki}A_{ij}C_{kj}= 0 \nonumber\\
 \Leftrightarrow
 &\forall k& [CAC^T]_{kk}= 0.
\end{eqnarray}
This set of {\em linear} equations may eliminate up to $N$ out of
$N(N-1)/2$ coefficients $A_{ij}$, thus
\begin{equation}
 \sfrac{N(N-1)}{2}\ge \dim({\NTS_a}^+) \ge \max\{\sfrac{N(N-3)}{2},0\}.
\end{equation}
The space $\NTS_b\equiv \Kern{\rho_b}\cap \Supp{\rho_a}$ on the other hand is
given by all vectors out of $\Supp{\rho_a}$, which are orthogonal to
$\ket{\psi_i\psi_j}+ \ket{\psi_j\psi_i}$ for all $i>j$. With a diagonal
coefficient matrix $B$ this yields
\begin{equation}
 \forall i>j\quad [CBC^T]_{ij}=0.
\end{equation}
Thus, we have
\begin{equation}\label{e18720}
 N\ge \dim \NTS_b\ge \max\{\sfrac{N(3-N)}{2},0\}.
\end{equation}
Since the dimension of the reduced Hilbert space is given as $\dim \Hilbert'=
\dim \Hilbert-(\dim{\NTS_a}^- +\dim{\NTS_a}^+)- \dim\NTS_b$, we finally arrive
at the main result of this section,
\begin{equation}
 0\le \dim \Hilbert' \le \left\{\begin{array}{ll}
  2 &\mbox{if}\quad N=2\\2N &\mbox{if}\quad N>2.
 \end{array}\right.
\end{equation}
The case $N= 2$, considered in section~\ref{s32551}, turns out to play a
special role, since here always $\dim \NTS_b>0$ holds, cf. (\ref{e18720}). We
point out that these bounds are tight. This can be directly verified by
considering a system of states with equal overlap, i.e. a system with
\begin{equation}\label{e4706}
 \quad \cos\vartheta\coloneqq \bracket{\psi_i}{\psi_j} \in [0,1[
  \quad \forall i\ne j.
\end{equation}
Then for the trivial case (i.e. $\vartheta= \sfrac \pi 2$) $\dim \Hilbert'=0$
holds, while the upper bound is reached whenever $\vartheta< \frac \pi 2$.
Thus state comparison for two out of three states may already lead to a
non-trivial UD problem, as illustrated in the following.

\subsection*{Example: ``two out of three''}
As an example of a case, where state comparison does not reduce to UD of pure
states, $N= 3$ is considered. We specialize to the case where the
states $\ket{\psi_1}$, $\ket{\psi_2}$ and $\ket{\psi_3}$ subject to comparison
satisfy (\ref{e4706}) with $0<\vartheta< \frac \pi 2$ and assume all {\em a
priori} probabilities to be equal, $q_1= q_2= q_3= \frac 1 3$.

The previous discussion of the related UD problem showed, that this related
problem can be reduced to a Hilbert space $\Hilbert'$ of dimension $\dim
\Hilbert'=\dim \Supp{\rho_a}+ \dim \Supp{\rho_b}= 3+3$. Since $N= 3$ this has
the consequence, that ${\NTS_a}^+= \NTS_b= \{0\}$. Thus, $\Hilbert'$ exactly
consists of the symmetric subspace of $\Hilbert\equiv \Supp{\rho_a}\oplus
\Supp{\rho_b}$, i.e.  $\Hilbert'=\Hilbert^+$. However, for the remaining UD
problem, no general optimal solution is known and we thus calculate the
tightest upper and lower bounds for the rate of success known so far, i.e. the
lower bound provided by Rudolph {\em et al.} \cite{Rudolph:2003PRA} and the
upper bound shown by Raynal and L\"utkenhaus \cite{Raynal:2005QPh}. These
bounds together with the rate of success for the separable measurement
are shown in \fig{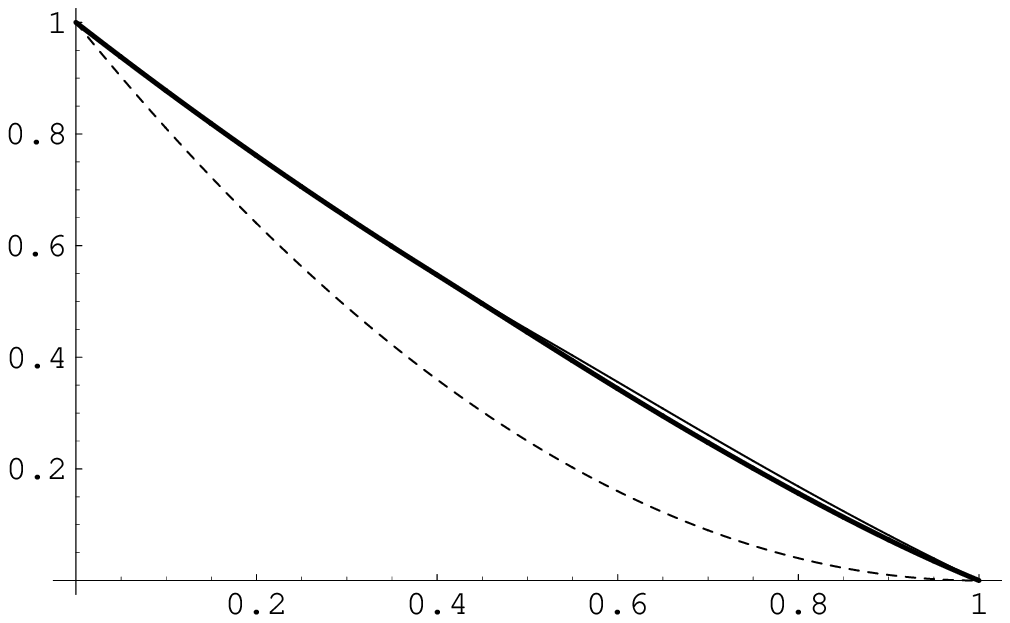}{0.61804}{
 \put(-1.04,0.34){\rotatebox{90.0}{\makebox(0,0){$P$}}}
 \put(-0.46,-0.02){\makebox(0,0){$\cos\vartheta$}}
}{Bounds for the probability of success for state comparison ``two out of
three'', with equal {\em a priori} probabilities and relative angles. The solid
lines are an upper \cite{Raynal:2005QPh} and a lower bound
\cite{Rudolph:2003PRA}, while the dashed line corresponds to the separable
measurement.}. Again, the incoherent measurement is always worse then the
measurement used to construct the lower bound. In addition one finds, that for
\begin{equation}
 \cos\vartheta \le \frac{\sqrt 2-\sqrt{\sqrt 2}}{2-\sqrt 2}
\end{equation}
(i.e. $\vartheta/\pi \apprge 0.375$) the lower and upper bound coincide,
revealing the {\em optimal} solution of UD of $\rho_a$ and $\rho_b$ in that
region to be
\begin{equation}
 P_\mathrm{opt}= 1-\frac{\sqrt{8}}{9}(4\cos\vartheta-\cos^2\vartheta).
\end{equation}
One can also show, that in this region the optimal measurement satisfies the
defining property~(\ref{e6923}) and thus also solves the problem of optimal
state comparison.

\section{Mixed State Comparison}\label{s19384}
In this section we investigate, in what situations a measurement can exist,
which satisfies the defining property~(\ref{e6923}). We have the following
\begin{proposition}
Unambiguous state comparison ``$C$ out of $N$'' for a set of mixed states
$\{\pi_1,\dotsc,\pi_N\}$ with arbitrary, but non-vanishing {\em a priori}
probabilities, can be realized iff $\forall \,i$
\begin{equation}\label{e295}
 \Supp{\pi_i} \nsubseteq \sum_{k \ne i} \Supp{\pi_k}.
\end{equation}
\end{proposition}
\paragraph*{Proof.}{
For the {\em if} part it is enough to show, that there is a POVM, given by
$\{\widetilde F_1, \dotsc, \widetilde F_N, \widetilde F_?\}$, such that
\begin{equation}\label{e28546}
 \tr(\widetilde F_i \pi_j)>0 \quad\Leftrightarrow\quad i=j.
\end{equation}
In order to construct such an POVM, denote by $\mathrm P_i$ the projector onto
the ortho-complement of $\sum_{k \ne i} \Supp{\pi_k}$. Then from equation
(\ref{e295}) it follows, that there is at least one vector $\ket{\varphi} \in
\Supp{\pi_i}$, such that $\ket{\phi_i}\coloneqq \mathrm P_i \ket{\varphi}$
satisfies $\bracket{\phi_i}{\phi_i}= 1$. These vectors $\ket{\phi_i}$ by
construction satisfy $\bra{\phi_i}\pi_i\ket{\phi_i}> 0$ for each $i$, while
$\bra{\phi_i}\pi_j\ket{\phi_i}= 0$ for all $j\ne i$. The choice
$\widetilde F_i=\sfrac 1 N \pr{\phi_i}$ satisfies (\ref{e28546}) and further
has $\widetilde F_?=\id - \sum_i \widetilde F_i \ge 0$. Indeed, for any $\ket
\psi$ out of the complete Hilbert space,
\begin{equation}
 \bra \psi \widetilde F_?\ket \psi = \bracket \psi \psi-
   \sfrac 1 N \sum_i|\bracket{\phi_i} \psi |^2 \ge 0
\end{equation}
holds by virtue of the Cauchy-Schwarz inequality.

For the {\em only if} part we use, that any unambiguous state comparison
measurement solves (not necessarily in an optimal way) the related unambiguous
state discrimination problem. However, assuming that for some $i$:
\begin{equation}\label{e21927}
 \Supp{\pi_i}\subset \sum_{k\ne i} \Supp{\pi_k},
\end{equation}
we show, that no UD measurement can satisfy $\tr(F_a {\pi_i}^{\otimes C})>0$,
thus being a contradiction to (\ref{e11971}).

In order to show this contradiction, note, that for positive operators $A$ and
$B$,
\begin{subequations}
\begin{eqnarray}
 \Supp{A+ B}&=& \Supp A+ \Supp B,\\
 \Supp{A\otimes B}&=& \Supp A \otimes \Supp B.
\end{eqnarray}
\end{subequations}
Further we use a Lemma, shown by Raynal, L\"utkenhaus and van Enk in
\cite{Raynal:2003PRA}, which states, that $\tr(AB)= 0$, iff $\Supp A \perp
\Supp B$. Now, assuming (\ref{e21927}), it follows that
\begin{equation}
 \Supp{{\pi_i}^{\otimes C}}= \Supp{\pi_i}^{\otimes C}\subset \sum_{k \ne i}
  \Supp{\pi_k}\otimes
 \Supp{\pi_i}^{\otimes (C-1)}\subset \Supp{\rho_b}.
\end{equation}
However, by the Lemma of \cite{Raynal:2003PRA}, the requirement $\tr(F_a
\rho_b)= 0$ (cf.  (\ref{e27918})) is equivalent to $\Supp{F_a}\perp
\Supp{\rho_b}$. This implies $\Supp{F_a}\perp \Supp{{\pi_i}^{\otimes C}}$ or
equivalently $\tr(F_a{\pi_i}^{\otimes C})=0$ and completes the proof.\qed }

For the comparison of qubits this proposition implies that unambiguous
comparison ``$C$ out of $N$'' can only be realized for $N=2$ and {\em pure
states}. For unambiguous state comparison ``$C$ out of $N$'' of pure states in
any dimension, Proposition 1 reduces to the result of Chefles {\em et al.}
\cite{Chefles:2004JPA}. They found that state comparison can only be realized
for linearly independent states. Another direct consequence from Proposition 1
is the fact that density matrices which contain a proportion of the identity
(e.g. by being sent through a depolarising channel, or by adding white noise in
an experiment) can never be compared unambiguously.

\section{Conclusions}\label{s29399}
We have addressed the question of unambiguous state comparison with general
{\em a priori} probabilities. Our method consists of reducing the corresponding
problem of unambiguous mixed state discrimination to a non-trivial subspace
\cite{Raynal:2003PRA}. We analytically solve the case for comparing two states
drawn from a set of two states, finding the optimal POVMs and the optimal rate
of success. There is a considerable gain of the optimal coherent strategy over
the best incoherent strategy. While this case reduces to the discrimination
between two pure states, the comparison of two states drawn from a set of three
states is shown to lead to a non-trivial mixed state discrimination task. So
far, the optimal solution is only found for certain parameter ranges.

The more general task of comparing two states from a set of $N$ states is
exceedingly difficult. No general solution to this problem exists. Here, we
have presented an upper bound for the dimension of the reduced Hilbert space.
This bound is shown to be reached for states with equal overlap. We have also
provided a necessary and sufficient condition for unambiguous comparison of
mixed states to be possible.

{\em Note added:} While completing this manuscript, we learned about related
work by Herzog and Bergou \cite{Herzog:2005QPh}, who found the same expression
as equation~(\ref{e26445}) for optimal unambiguous state comparison of two
states drawn from a set of two states.

\acknowledgments{We would like to thank N. L\"utkenhaus, T. Meyer, and P.
Raynal for enlightening discussions. Furthermore we would like to thank the
anonymous referee for very valuable comments. This work was partially supported
by the EC programme SECOQC.}

\appendix
\section{Optimal separable measurement ``two out of two''}\label{a348}
This appendix is dedicated to show that with the na\"{i}ve measurement given in
equation (\ref{e20301}), indeed the optimal separable solution was found. That
is, the optimal {\em separable} unambiguous state comparison measurement for
two states drawn from a set of two pure states $\{\ket{\psi_1}, \ket{\psi_2}\}$
is solved in an optimal way by performing optimal unambiguous state
discrimination in each subsystem.

A general element of a separable POVM $\{F_x\}$ is of the form
\begin{equation}
 F_x= \sum_{i,j} c_{x,ij} F_{x,i}^{(1)}\otimes F_{x,j}^{(2)},
\end{equation}
where the non-negative coefficients $c_{x,ij}$ account for the relative
contribution of each of the terms containing the positive local POVM elements
$F^{(k)}_{x,i}$.

First we show, that in our case no measurement outcome of either subsystem can
be used to adapt the measurement of the other. Consider without loss of
generality, that a measurement first takes place in subsystem~$1$ and yields
with probability $p^{(1)}_{x,i}$ the outcome $(x,i)$. This measurement is
applied to the global state $\rho\coloneqq \eta_a \rho_a +\eta_b \rho_b= (q_1
\pr{\psi_1} + q_2 \pr{\psi_2})^{\otimes2}$ and yields in subsystem~$2$
\begin{equation}
 \tr_1((F_{x,i}^{(1)}\otimes \id) \rho)= p^{(1)}_{x,i}
  (q_1 \pr{\psi_1} + q_2 \pr{\psi_2}),
\end{equation}
which is, up to the factor $p^{(1)}_{x,i}$ independent of the outcome $(x,i)$.
Thus the local measurements can be optimized in each subsystem separately, and
one is free to choose the same (optimal) measurement in both systems due to the
symmetry of $\rho_a$ and $\rho_b$. Therefore we can drop the upper label $(k)$
on the local measurement elements in the following.

Furthermore one is forced to choose these measurements to be UD measurements.
Indeed, $\tr(F_a \rho_b)= \tr(F_b \rho_a)= 0$, only if for each $x\in \{a,b\}$
and for all $l$, either $\tr(F_{x,l} \pr{\psi_1})= 0$ or  $\tr(F_{x,l}
\pr{\psi_2})= 0$. We prove this statement by contradiction: Suppose, that at
least one term ($c_{a,ij} F_{a,i}\otimes F_{a,j}$) of $F_a$ contains at
least one local POVM element $F_{a, m}$ (where $m\in \{i,j\}$), having a
non-vanishing expectation value for both states, i.e.
\begin{equation}
 \bra{\psi_1}F_{a,m} \ket{\psi_1} >0 \mx{and}
 \bra{\psi_2}F_{a,m} \ket{\psi_2} >0.
\end{equation}
It follows that
\begin{equation}
 \tr((c_{x,ij} F_{a,i}\otimes F_{a,j})\rho_a) >0
\end{equation}
and
\begin{equation}
 \tr((c_{x,ij} F_{a,i}\otimes F_{a,j})\rho_b) >0,
\end{equation}
which is in contradiction to $\tr(F_a \rho_b)= 0$. An analogous argument holds
$F_b$.

Without loosing any information, a UD measurement can always be reduced to
have the measurement elements $\{F_1,F_2,F_?\}$, with $\bra{\psi_2} F_1
\ket{\psi_2}= \bra{\psi_1} F_2 \ket{\psi_1}= 0$.  In order to make this a valid
choice for the local measurements of unambiguous state comparison, in addition
the conditions (\ref{e6923}) have to be satisfied, i.e.
\begin{equation}
 \alpha\coloneqq \bra{\psi_1}F_1\ket{\psi_1} > 0\mx{and}
 \beta\coloneqq \bra{\psi_2}F_2\ket{\psi_2} > 0.
\end{equation}

From the consideration above, we find that $F_a$ and $F_b$ are of the form
\begin{eqnarray}
 F_a&=& F_1 \otimes F_1+ F_2 \otimes F_2, \\
 F_b&=& F_1 \otimes F_2+ F_2 \otimes F_1.
\end{eqnarray}
The optimal separable state comparison corresponds to $F_1= \widetilde F_1$ and
$F_2= \widetilde F_2$ as above defined (\ref{e20301}). Thus we have shown, that
in this case the optimal separable unambiguous state comparison strategy is
indeed given by consecutive optimal UD measurements.

Let us mention that for the optimal UD measurement the conditions $\alpha >0$
and $\beta >0$ do not always hold: in those situations, where
condition~(\ref{e31616}) is not satisfied, $\alpha= 0$ or $\beta= 0$. But
changing $\alpha$ and $\beta$ (under the constraint $\id- F_1 -F_2\ge 0$)
infinitesimally, affects the probability of success only infinitesimally. In
this limit, we consider the optimal unambiguous state discrimination
measurement as a valid choice for $F_1$ and $F_2$.

We conjecture that also in the more general scenario of unambiguous state
comparison of ``C out of N'' states, the best separable measurement is given by
performing unambiguous state discrimination in each subsystem. However, the
proof by contradiction given above for ``two out of two'' cannot be
generalized in a straightforward way for the operator $F_b$. We leave the
generalization as an open question for future work.

\bibliography{thebib}
\end{document}